\documentclass[]{aa501}
\usepackage{times,graphicx}
\usepackage{epsfig}
\begin{document}

\title{More nitrogen rich B-type stars in the SMC cluster, NGC 330}

\subtitle{}

\author{D.J.~Lennon\inst{1} \and P.L.~Dufton\inst{2}
\and C.~Crowley\inst{1}}

\offprints{D.J.~Lennon}

\institute{Isaac Newton Group of Telescopes, Apartado de Correos 321, E-38700, 
Santa Cruz de La Palma, Canary Islands, Spain \newline
email: djl@ing.iac.es
\and
The Department of Pure and Applied Physics, The Queen's University
of Belfast, Belfast BT7 1NN, N.~Ireland \newline
email: P.Dufton@Queens-Belfast.ac.uk
}

\date{Received date; accepted date}

\abstract{
High resolution spectra of seven early B-type giant/supergiant stars 
in the SMC cluster NGC330 are analysed to 
obtain their chemical compositions relative to SMC field and 
Galactic B-type stars. It is found that all seven stars 
are nitrogen rich with an abundance approximately 1.3 dex 
higher than an SMC main-sequence field B-type star, AV304.
They also display evidence for deficiencies in carbon,
but other metals have abundances typical of the SMC.
Given the number of B-type stars with low rotational projected
velocities  in NGC330 (all our targets have $v$sin$i < 50$\,km/s),
we suggest that it is unlikely that the stars in our sample 
are seen almost pole-on, but rather that they are intrinsically 
slow rotators. Furthermore, none of our objects displays any evidence of 
significant Balmer emission excluding the possibility that these 
are Be stars observed pole-on. Comparing these results with 
the predictions of stellar evolution models including the effects 
of rotationally induced mixing, we conclude that
while the abundance patterns may indeed be
reproduced by these models, serious discrepancies
exist. Most importantly, models including the effects of initially 
large rotational velocities do not reproduce the
observed range of effective temperatures of our sample, nor the
currently observed rotational velocities.
Binary models may be able to produce stars in the observed
temperature range but again may be incapable
of producing suitable analogues with low rotational velocities.  
We also discuss the clear need for stellar evolution calculations 
employing the correct chemical mix of carbon, nitrogen and oxygen for 
the SMC.
}

\maketitle

\section{Introduction} 

NGC330 is one of the brightest and most populous
young clusters in the Small Magellanic Cloud (SMC).
The photometric surveys of 
Arp (\cite{Arp}), Robertson (\cite{Rob74}) and  
Carney et al. (\cite{Car}) illustrate the
key features of the cluster's
colour-magnitude diagram, namely the presence of two
groups of blue and red supergiants well separated from the 
cluster's supposed main-sequence blue plume.  These two groups of
stars have been widely interpreted 
as core helium burning stars and
the cluster is therefore considered as a key test
of stellar evolution theory and physics for stars of intermediate mass
in a low metallicity regime.  Essentially the ratio of
blue (B) to red (R) supergiants is an indicator of the
relative times a massive star spends in the these
phases, and these quantities are extremely sensitive to
assumptions made concerning convection and mixing. 
In fact the B/R ratio in NGC330 is generally assumed to
be representative of the SMC as a whole and is used
as a calibrator for stellar evolution calculations at
low metallicity (Stothers \& Chin \cite{SCa}, \cite{SCb}; 
Keller et al. \cite{Kel00}; Chiosi et al. \cite{Chi95}).  
The specific problem of the B/R ratio as a function of
metallicity has been discussed by Langer \& Maeder (\cite{Lan95}),
where a more detailed discussion of the various treatments
of convection and overshooting may be found. 

The interpretation of the cluster's HR diagram is complicated by the
surprise finding that many main sequence B-type stars in the cluster
have H$\alpha$ emission implying a very high incidence of Be stars 
(Feast \cite{Fst72}). A subsequent 
intermediate band and H$\alpha$ photometric study
indicated that at least 60\% of all main-sequence B-type stars are of 
Be-type (Grebel et al. \cite{Gre96}), this high fraction being 
confirmed independently by the spectroscopic
observations of Lennon et al. (\cite{Len94}), Mazzali et al
(\cite{Maz96}) and Keller \& Bessell (\cite{Kel98}).  
As with the ratio of B/R supergiants,
the ratio Be/B-type stars in NGC330 is often taken as being representive
for the SMC metallicity (Maeder et al. \cite{Mae99}).
A second complication arises concerning
uncertainty over the metallicity of stars in NGC330;
some estimates of the metallicity based upon spectroscopy
of the brightest K and F-type supergiants (Spite et al. \cite{Spi91}) 
and one B-type giant
(Reitermann et al. \cite{Rei90}) imply that these objects are metal poor
even with respect to SMC field stars, while Str\"omgren photometric
observations of supergiants by 
Grebel \& Richtler (\cite{Gre92}) have been interpreted as
evidence for a metal deficiency of 0.5 dex with respect to field stars.
However more recent analyses of K-type supergiants have tended to
suggest that this difference in metallicity is much smaller,
or indeed not significant (Hill \cite{Hil99}), confirming
the results obtained from the analysis of two B-type stars
in the cluster by Lennon et al. (\cite{Len96}, hereafter Paper I).

The spectroscopic work of Lennon et al. (\cite{Len94}) also
found that the bright non-Be and weak Be-type stars occupied that 
region of the HR-diagram known as the post main sequence gap,
or blue Hertzsprung gap (BHG).
That is, they are giant/supergiant stars lying red-wards of the
main sequence band, but blue-wards of the A/F-type supergiant regime.
Caloi et al. (\cite{Cal93}) and
Grebel et al. (\cite{Gre96}) have also commented on this fact, the latter
suggesting that these stars are most likely a mixture of rapidly rotating
B/Be-type stars of varying orientation and blue stragglers formed by
interaction in binary stars.  Keller et al. (\cite{Kel00}) also attempted
to address this problem using far-UV photometry
(from the F160BW filter on WFPC2 of HST) to constrain
B-type stellar effective temperatures and find significantly fewer
stars in the blue Hertzsprung gap (BHG). However they
assumed that the logarithmic surface gravities were
4.0 in their work (Keller, private communication), which may result
in spuriously high effective temperatures for stars
near the turn-off since they have much lower surface gravities
(Lennon et al. \cite{Len94}). Note that Caloi
et al. (\cite{Cal93}) also adopted lower values for the
surface gravities.  Clearly a detailed spectroscopic analysis
of the BHG stars leading to estimates of both stellar parameters 
and atmospheric abundances is needed for comparison with 
the predictions of various stellar
evolution calculations.

In Paper I, we derived metallicities of two such 
B-type stars in NGC330 and while we found them in general
to be compatible with that of the SMC field both stars had
a significant nitrogen overabundance. The magnitude of 
the nitrogen enrichment was uncertain due to the small number of 
NGC330 targets analysed and also the difficulty in estimating 
the low nitrogen abundance of the SMC field, at least, from
B-type stars.  Also in Paper I we found that the carbon
abundance was not significantly depleted, contrary to what one
expects if the nitrogen were produced in the CN cycle.
The carbon abundance was uncertain however, and coupled
with the uncertainty of the magnitude of the nitrogen
enrichment, made interpretation difficult.
An additional puzzling aspect was that both stars are
narrow-lined and therefore if the nitrogen enhancements
are the result of high rotation we must be observing them
almost pole-on, which seemed unlikely give that they
were drawn from the sample of about 20 stars observed
by Lennon et al. (\cite{Len94}).  In the present paper 
we analyse seven targets in NGC330 (including 
the two discussed in Paper I) belonging to both the
blue supergiant group and the tip
of the blue main sequence plume discussed above. 
Hence all stars lie in, or close to, the BHG.
We estimate stellar parameters, radial and projected rotational
velocities, as well as both absolute
abundances and differential abundances relative to a galactic
target in the h and $\chi$ Persei cluster (Vrancken et al. 
\cite{Vra00}) and to an SMC field star, AV304. For the latter
we utilise the results from a recent analysis (Rolleston et al
\cite{Rol02}) based on high quality VLT data, which now
give us a reliable estimate of the pristine nitrogen 
abundance in unevolved B-type stars in the SMC.  
We also attempt to provide improved estimates for
carbon abundances and compare our
results with stellar evolution calculations, including
the recent models of Maeder \& Meynet (\cite{Mae01}) which include
the effects of rotationally induced mixing.

\section{Observational data and results}
\subsection{Optical Spectroscopy}

High resolution spectra of early-type stars in the cluster NGC330, 
were obtained using the CASPEC spectrograph on the ESO 3.6m on the
26 and 27 August 1994 and on the 22 and 23 September 1996; 
the observational parameters are summarized in Table \ref{obsdata}.
The UBV photometry has been taken from Mazzali et al. (\cite{Maz96})
supplemented by additional unpublished data, while the spectral types 
have been estimated from unpublished low dispersion EMMI spectroscopy.  
The observational configuration and data reduction techniques are 
discussed in detail in Paper I.
Briefly, Caspec was used with an entrance slit width of 
2 arc-seconds giving an effective resolving power of
approximately 20\,000, or about 15\, km/s. 
Preliminary reduction of the echellograms to a one dimensional 
format was achieved using the IRAF reduction package (Willmarth 
and Barnes, \cite{Wil94}). The relative faintness of the 
targets, lead to some of the spectra having relatively
low continuum counts and signal-to-noise ratios. 
In particular, the observations for A01, B04 and B32 were
taken in conditions that were at times cloudy, and so not
all spectra were co-added to produce the final spectrum for these
stars. The s/n estimates 
are summarized in Table \ref{obsdata} and were measured near
the centre of the echelle order at approximately 4200\AA.
Fortunately, all the stars have small projected rotational velocities
and are hence sharp lined with line widths (FWHM) ranging from 
0.4 to 0.9\AA~(see Table \ref{obsdata}). Hence despite the 
significant shot noise, lines with equivalent widths of more than 
50m\AA~could normally be detected, while features with line strengths
as small as 30m\AA~could sometimes be convincingly identified.
Projected rotational velocities were estimated from the
line widths by assuming that the only contributions to the 
intrinsic widths of the absorption lines are 
instrumental and pure rotational i.e. that the effects of 
macroturbulence are negligible. Hence  the results in Table \ref{obsdata}
are most likely upper limits on the projected rotational velocities.

Equivalent widths were measured for the metal and non-diffuse helium
lines using the {\sc starlink} spectrum analysis program {\sc dipso}. 
Absorption lines were fitted using a non-linear least squares 
technique with a variety of profile shapes (e.g. triangles, gaussians) 
and degrees of polynomial for the continuum being considered. 
Equivalent widths for the diffuse helium lines were measured manually
with the continuum being arbitrarily defined at $\pm$10\AA~
from the line centre. For the hydrogen H$\gamma$ and H$\delta$ 
lines, the spectra were again normalised and profiles measured, 
with the continuum levels now being defined at $\pm$16\AA~(as 
were the theoretical profiles). For all these measurements the 
procedures were effectively identical to those described in Paper I
where further details can be found.

The model atmosphere analysis utilised both absolute and 
differential techniques. For the latter, two standard stars were 
considered - BD +56\,576 in the h and $\chi$ Persei galactic 
cluster and the SMC target AV304. The observational data 
for BD +56\,576 was taken from our study (Vrancken et al
\cite{Vra00}) of early-B type stars in this double cluster,
while the results for AV304 are based on recently obtained
VLT spectra (Rolleston et al. \cite{Rol02}).

\begin{table*}
\begin{flushleft}
\caption[]{Observational data for NGC330 targets: Star identifications
are from Robertson (\cite{Rob74}). Spectral types are from
Lennon et al. (\cite{Len94}) or Grebel et al. (\cite{Gre96})
although some classifications are slightly revised.
Photometry is taken mostly
from Mazzali et al.\, (\cite{Maz96}) with additional data taken
from Grebel et al.\, (\cite{Gre96}) and Robertson (\cite{Rob74}).
As discussed in the text, the projected rotational velocities
($v$sin$i$) are most probably upper limits, and are accurate
to $\pm5$\,km/s. Radial velocities ($v_{rad}$) are corrected
to heliocentric values.
}
\label{obsdata}
\begin{tabular}{llcrrrrrccc} 
  \hline
  \hline 
Star  & Sp. Type   & V     & B-V    & U-B    & Date & Exposures & 
s/n 	 & FWHM & $v$sini & $v_{rad}$\\
  
      &            &       &        &        & yymmdd & n x secs & 
 & \AA & km/s & km/s \\
\hline
\\
A01   & B0.5\,III/V  & 14.70 & $-$0.18  & $-$0.85  & 940828 & 3x3600 &
40	 & 0.6 & 30 & $+151\pm 6$ \\
\\
A02   & B4\,Ib       & 12.90 & $-$0.05  & $-$0.69  & 940828 & 2x1200 &
69	 & 0.6 & 30 & $+151\pm 3$ \\
&&&&&& 1x600 &&&& \\
\\
A04   & B2-3\,IIe  & 14.54 & $-$0.03 & -- & 960923 & 2x3000 &
27	 & 0.4 & 20 & $+155\pm 5$ \\
\\
B04   & B2\,III      & 15.60 & $-$0.16  & $-$0.82  & 940827 & 3x3600 &   
25	 & 0.4 & 20 & $+150\pm 4$ \\
&&&&&& 2x1800 &&&& \\
\\
B22   & B2\,II       & 14.29 & $-$0.13  & $-$0.75  & 940828 & 1x3600 &
51	 & 0.8 & 35 & $+157\pm 4$ \\
&&&&&& 1x2100 &&&& \\
\\
B32   & B2\,III & 15.01 & $-$0.17 & -- & 960923 & 2x3600 &
22	    & 0.9 & 40 & $+157\pm 7$ \\
&&&&& 960924 & 2x3600 &&&& \\
\\
B37   & B3\,Ib       & 13.19 & $-$0.07  & $-$0.80  & 940827 & 2x1800 &  
90	 & 0.8 & 35 & $+156\pm 3$ \\
\\
  \hline   
\end{tabular}
\end{flushleft}
\end{table*}

\subsection{Flux distributions}

IUE low resolution spectra are available in both the short
and long wavelength cameras for all the NGC330 targets 
(apart from A04) and for BD +56\,576. These were  extracted 
from the INES archive (Rodriguez-Pascuel et al. \cite{Rod99}), 
where further details may be found.

\section{Method of Analysis}
We initially consider standard LTE model atmosphere techniques 
in our analysis, however in section 5 we discuss the accuracy
of this method and correct our abundances for NLTE effects.
Here we use the
model atmospheres taken from the grids calculated with the
code of Kurucz (\cite{Kur91}) and made available at his
website. This grid covers a range of chemical compositions, 
with the current analysis utilising mainly the results for 
metallicities of $-$0.5 dex, which is compatible
with that found in Paper I. Details of the methods and atomic data 
used in the radiative transfer calculations can be found in, for 
example, Rolleston et al. (\cite{Rol00}) or Smartt et al. (\cite{Sma96}).

\subsection{Atmospheric parameters}
As discussed in Paper I, the relatively low signal-to-noise
of the Caspec spectra preclude the observation of different 
ionization stages of the same element for most of the stars. 
Additionally there is no extant Str\"omgren photometry. 
Hence none of the standard techniques (see, for example, Kilian,  
\cite{Kil94}, Gies and Lambert, \cite{Gie92}, Rolleston et al.
\cite{Rol00}) are available for constraining effective
temperatures. Hence, initial temperature estimates were made from the
observed flux distributions. The extinction toward NGC330 has been
found to be relatively small and here we have adopted a foreground
reddening of $E(B-V)=0.034$ and the extinction relation of
Seaton (\cite{Sea79}), together with an SMC extinction of $E(B-V)=0.05$
and the relationship of Thompson et al. (\cite{Tho88}). As discussed
in Paper I, similar effective temperatures would have been estimated
using, for example, the galactic extinction law for all the
reddening.

The effective temperature estimates are listed in Table \ref{abs_analysis}.
As discussed in Paper I, at effective temperatures greater than
approximately 22\,000~K, the flux distribution becomes relatively
insensitive to this parameter. Additionally there may be uncertainties
due to the presence of fainter targets in the IUE aperture. Hence
we adopt conservative error estimates of 2\,000~K for our cooler targets 
and up to 4\,000~K for our hottest targets. For the latter, we note
(see Paper I for details) that our estimates are consistent with
the absence of a detectable \ion{He}{ii} spectrum, which implies
that the effective temperatures must be less than 26\,000~K.

\begin{table*}
\caption{Atmospheric parameters (effective temperatures (K),
logarithmic surface gravities (cgs), microturbulent velocities (km/s))
and LTE abundances for NGC330 targets (the numbers in parentheses
give the number of lines used to derive the mean abundance).
For stars without an error estimate for the microturbulent
velocity we have assumed a value of 5 km/s.
}
\label{abs_analysis}
\begin{center}
\begin{tabular}{lllllllll}
  \hline   
  \hline
Star &{A01~~~~} &{A02~~~~} &{A04~~~~} &{B04~~~~} &{B22~~~~} &{B32~~~~} &{B37~~~~} & Mean \\
  \hline
T$_{\rm eff}$ & 24\,000 & 16\,000 & 18\,000    & 23\,000 & 20\,000 & 22\,000 & 18\,000 \\
log g         & 3.7$\pm$0.3 & 2.3$\pm$0.1 & 2.8$\pm$0.2 & 3.6$\pm$0.2 & 
3.0$\pm$0.2 & 3.0$\pm$0.3 & 2.4$\pm$0.1 \\
v$_{\rm t}$   & 5       & 8$\pm$4 & 5    & 5       & 11$\pm$4 & 5       & 10$\pm$4 \\
\\
He I  &10.78~(10)&10.58~(10)&10.66~(9) &10.78~(10)&10.82~(10)&10.85~(8)&10.98~(10)&10.78$\pm$0.13\\
C II  &6.82~(1)  &6.91~(2)  & 7.05~(2) &6.86~(1)  &6.89~(2)  &6.54~(1) &6.71~(2)  &6.85$\pm$0.15	   \\
N II  &7.62~(8)  &7.71~(9)  & 7.16~(2) &7.54~(7)  &7.52~(7)  &7.40~(3) &7.89~(9)  &7.62$\pm$0.18       \\
O II  &8.13~(16) &8.14~(12) & 8.23~(2) &7.88~(5)  &7.83~(7)  &8.23~(3) &7.98~(12) &8.05$\pm$0.13	   \\
Mg II &6.78~(1)  &6.69~(1)  & 6.43~(1) &6.68~(1)  &6.54~(1)  &--       &6.79~(1)  &6.65$\pm$0.14	   \\
Si II &--        &6.65~(2)  & 6.26~(2) &--        &--	     &--       &	  &6.5:   \\
Si III&6.75~(3)  &6.82~(2)  & 6.08~(2) &6.55~(3)  &6.49~(3)  &6.25~(1) &7.12~(3)  &6.63$\pm$0.32	   \\
Si IV &7.19~(1)  &-- 	    & --       &--        &--	     &--       &--	  &7.2:	   \\
\hline
\end{tabular}
\end{center}
\end{table*}

Two stars, A01 and B04, had been already been discussed in Paper I; here
we have independently re-estimated their effective temperatures. For the latter
the two estimates are consistent, whilst for A01, the current estimate
is 2\,000~K less than in Paper I. Here, we used flux distributions calculated
for a metallicity 0.5 dex less than solar (in paper I, fluxes for a normal
metallicity were adopted) but tests showed that the estimates were little
affected by the choice of metallicity; hence we ascribe the difference mainly
to the flux distributions not being particularly temperature sensitive
in the case of A01.

For A04, no IUE observations were available; there are, however, measurements
of the equivalent widths of both \ion{Si}{iii} and \ion{Si}{iv} ions and
these have been used to estimate an effective temperature (see, for example,
Rolleston et al. \cite{Rol00} for details) and this is also listed in Table
\ref{abs_analysis}. Note that the equivalent widths are small (typically
30 m\AA) and hence subject to considerable uncertainty. Additionally
several authors (for example, Kilian \cite{Kil92}, Smartt et al.
\cite{Sma01}) have found systematic differences between temperatures
found from optical photometry and ionization equilibria and it is
possible that similar differences will be present for the two
methods used here. Hence, the temperature estimate for A04 should be
treated with some caution.

Comparison of observed and theoretical Balmer lines (see Rolleston
et al. \cite{Rol00} for further details) were then used to
estimate logarithmic surface gravities, which are again summarized
in Table \ref{abs_analysis}. Some of the surface gravity estimates
differed significantly from the value (3.5 dex) assumed when
initially estimating the effective temperatures.  In these cases
the effective temperatures were re-estimated using an appropriate
gravity and the procedure iterated until it converged.
An example of the quality of the fit can be
found in Fig.~2 of Paper I and relevant uncertainties for the
individual stars are also listed in Table \ref{abs_analysis}.

For three stars (A02, B22, B37) there were sufficient \ion{O}{ii} lines 
to estimate microturbulent velocities and these are again summarized
in Table \ref{abs_analysis}. For the other stars a value of 5 km s$^{-1}$ was
adopted; we note that although this may be too low, the weakness of
the metal line spectra in the SMC targets implies that the choice
of this parameter does not significantly affect the abundance analysis.

For the galactic standard, BD+56\,576, the atmospheric parameters have
been re-estimated using the same criteria and methods as for the NGC330
targets and are summarized Table \ref{gal_stand}, together with the values
deduced by Vrancken et al. (\cite{Vra00}, VDLL). Given the different criteria
and non-LTE methods used by VDLL, the agreement is surprisingly
good. However, we emphasize that the values
deduced here should not be considered as the best available but rather
appropriate for a differential analysis with respect to the NGC330 targets. 
The comparison with VDLL also provides us with an estimate
of the importance of non-LTE effects which will be discussed in the 
next section.

In the case of AV304, we have adopted the atmospheric parameters
(listed in Table \ref{gal_stand}) deduced by Rolleston et al. 
(\cite{Rol02}). They used similar model atmosphere techniques and 
criteria to those adopted here, apart from the effective temperature
which was estimated from the silicon ionization equilibria. However,
in a previously analysis of this star (Rolleston et al. \cite{Rol93}) good
agreement was found between the effective temperatures estimates found
from the ionization equilibrium and the IUE flux distribution.

\begin{table}
\caption{Model atmosphere analysis of galactic standard, BD +56\,576,
together with the previous results of Vrancken et al. (\cite{Vra00}
- VDLL). The results for AV304 are taken directly from Rolleston et al. 
(\cite{Rol02}) Numbers in parentheses refer to the number of lines
used to derive the abundance.
}
\label{gal_stand}
\begin{center}
\begin{tabular}{lllllllll}
  \hline   
  \hline
       & {This paper} & VDLL   		& AV304	\\
Method & {LTE}        & non-LTE 	& LTE	\\
  \hline
T$_{\rm eff}$ & 22\,000 & 22\,500 	& 27\,500    \\
log g         & 3.4     & 3.4 		& 3.8        \\
v$_{\rm t}$   & 15      & 13		& 5          \\
\\
He    & 10.85~(7) & -	  &  10.86     \\
C     & 7.53~(2)  & 7.89  &  7.20 \\
N     & 7.41~(7)  & 7.35  &  6.66 \\
O     & 8.66~(16) & 8.57  &  8.23 \\
Mg    & 7.13~(1)  & 7.08  &  6.79 \\
Si    & 7.04~(6)  & 7.09  &  6.79 \\
\hline
\end{tabular}
\end{center}
\end{table}

\section{Results}

Using the atmospheric parameters discussed above, the metal lines 
in the NGC330 targets were used to deduce absolute abundances using
atomic data taken from Jeffery (\cite{Jef91}). These absolute values 
are listed in Table \ref{abs_analysis}, together with the number
of lines  used in the analysis. Also listed for each
ionization stage are the mean abundance estimate and the standard
deviation. In Table \ref{gal_stand}, absolute abundance estimates
are given for BD +56\,576 (both deduced here and by Vrancken et al.
\cite{Vra00}) and for AV304, which are again taken directly from
Rolleston et al. (\cite{Rol02}). 

\begin{table*}
\caption{Differential abundance analyses of the NGC330 targets and
of AV304 relative to galactic standards BD +56 576 and HR2387 
respectively. The number of lines (n) used in the derivation
of these values is also given.}
\label{diff_analysis}
\begin{tabular}{llrlrlrlrlrlr}
  \hline   
  \hline
Star &\multicolumn{2}{c}{He I} 
     &\multicolumn{2}{c}{C II}
     &\multicolumn{2}{c}{N II} 
     &\multicolumn{2}{c}{O II} 
     &\multicolumn{2}{c}{Mg II}
     &\multicolumn{2}{c}{Si III} \\
     & ~~~~$\Delta$[$\frac{He}{H}$] & n 
     & ~~~~$\Delta$[$\frac{C}{H}$] & n
     & ~~~~$\Delta$[$\frac{N}{H}$] & n
     & ~~~~$\Delta$[$\frac{O}{H}$] & n
     & ~~~~$\Delta$[$\frac{Mg}{H}$] & n
     & ~~~~$\Delta$[$\frac{Si}{H}$] & n
      \\
  \hline
A01  &--0.01$\pm$0.08 & 7  &--0.43          & 1  & +0.17$\pm$0.24 & 7 &  --0.53$\pm$0.24 & 16
& --0.35 & 1 & --0.46$\pm$0.30 & 3		\\
A02  &--0.31$\pm$0.17& 7  &--0.62$\pm$0.21 & 2  & +0.30$\pm$0.19 & 6 &  --0.54$\pm$0.30 & 12
& --0.44 & 1 & --0.37$\pm$0.06 & 2		\\
A04  &--0.19$\pm$0.18 & 6  &--0.49$\pm$0.30 & 2  &--0.19$\pm$0.21 & 2 &  --0.44$\pm$0.18 &  2
& --0.70 & 1 & --1.12$\pm$0.13 & 2    			\\
B04  &--0.02$\pm$0.12 & 7  &--0.39          & 1  & +0.13$\pm$0.33 & 7 &  --0.62$\pm$0.23 &  5
& --0.45 & 1 & --0.67$\pm$0.41 & 3		\\
B22  &+0.07$\pm$0.12  & 7  &--0.64$\pm$0.17 & 2  & +0.06$\pm$0.18 & 5 &  --0.85$\pm$0.39 &  7
& --0.59 & 1 & --0.73$\pm$0.23 & 3		\\
B32  &--0.01$\pm$0.14 & 5  &--0.71          & 1  &--0.09$\pm$0.03 & 2 &  --0.39$\pm$0.04 &  3
& --  & --    & --0.94          & 1		\\
B37  &+0.17$\pm$0.11 & 7  &--0.83$\pm$0.25 & 2  & +0.39$\pm$0.27 & 6 &  --0.64$\pm$0.21 & 12
& --0.34 & 1 & --0.09$\pm$0.22 & 3		\\
\\
Mean &--0.04$\pm$0.15 & -- &--0.61$\pm$0.15 & -- & +0.17$\pm$0.16 & -- &  --0.59$\pm$0.11 & --
& --0.48$\pm$0.14 & -- & --0.57$\pm$0.33 & --			\\
\\
AV304 &--0.01$\pm$0.04& 3 & --0.58$\pm$0.11 & 2 & --1.17$\pm$0.08 &2 
& --0.38$\pm$0.11 & 68 & --0.52 
& 1 & --0.50$\pm$0.03 & 2\\
\hline
\end{tabular}
\end{table*}

It is  more useful to consider a differential analysis
of the NGC330 targets relative to the galactic standard BD 56\,576. 
This will be less affected by any errors in oscillator strengths 
or systematic errors in the adopted atmospheric parameters. Unfortunately, 
it was not possible to consider all the lines observed in the Caspec spectra 
due to the more limited wavelength coverage for BD +56\,576. The differential 
abundances are summarized in Table \ref{diff_analysis} for each star, 
together with
the means (and standard deviations) for each element.  The results of
a similar differential comparison 
for AV304 (relative to the galactic star HR2387)
taken from Rolleston et al. (\cite{Rol02}) are also summarized in 
table \ref{diff_analysis}. These differential analyses confirm the
main results from the absolute analyses -- all the SMC targets are
deficient in metals apart from nitrogen, which is very deficient in
AV304 but slightly enhanced relative to our Galaxy in the NGC330 targets.
Below the results for individual elements are discussed in more detail.

\subsection{Helium} 
An effectively normal helium abundance is found, although there
is considerable scatter within the NGC330 stars. This probably
reflects the difficulty in estimating the continuum placement for
the relatively broad diffuse helium lines, which make up the
majority of our dataset. Hence we conclude that within the
uncertainties, there is no evidence for an anomalous helium
abundance in our targets.

\subsection{Carbon}  
The differential abundances are based mainly on the \ion{C}{ii} doublet 
at 4267\AA, supplemented in some cases by the line at 3921\AA. 
The former is know to be particularly sensitive to non-LTE effects
(Eber and Butler, \cite{Ebe88}). Also as discussed in Paper I
the stellar absorption line could be affected by recombination 
from an associated H{\sc ii} region. However all the NGC330 targets 
show a significant carbon underabundance (with a relatively small
scatter between stars), while the mean differential abundance is
similar to that found for AV304. 

The latter result appears inconsistent with the results in 
Tables 2 and 3, where the corresponding abundances differ by 
0.35 dex. However the value for AV304 was based  on a mixture of
\ion{C}{ii} and \ion{C}{iii} lines and if only the former are
considered the difference in absolute abundances decreases to
approximately 0.15 dex. The small remaining discrepancy of 
approximately 0.1 dex with the differential abundances arises 
from the different set of lines used in the two analyses.
 
\subsection{Nitrogen} 
As discussed in Paper I, the most striking difference between the two 
NGC330 targets and AV304 is in their \ion{N}{ii} spectra. However the new 
VLT observations for AV304, now allows this difference to be quantified.
Indeed the mean differential abundance implies that for the NGC330
targets nitrogen is
overabundant by a factor of approximately twenty compared with AV304. This
is consistent with paper I, where a lower limit on the 
nitrogen enrichment of 
a factor of six was deduced relative to the SMC. There is 
considerable scatter in the differential abundances deduced
for individual stars. This is somewhat surprising as the
\ion{N}{ii} lines are relatively strong and hence the
equivalent width estimates should be reliable. Additionally, 
only small non-LTE effects have been found for the \ion{N}{ii}
spectra (Becker and Butler, \cite{Bec89}). Hence the
scatter could possibly reflect a real variation in the
nitrogen abundances in these stars.

\subsection{Oxygen} 
The \ion{O}{ii} spectrum, together with that for \ion{N}{ii} is the
best observed in our targets. The differential oxygen abundances are
in good agreement with only that for B22 being far from the mean. 
This star has a relatively low effective temperature estimate of 
20\,000 K and a change of 1\,000 K in this estimate would change
the differential abundance by more than 
0.3 dex. Hence we believe that there is no evidence for any
variation in the oxygen abundances among the NGC330 targets.
Additionally the mean differential abundance is similar to,
although somewhat lower than, that found for AV304.
 
\subsection{Magnesium} 
Although the results are based solely on the 
\ion{Mg}{ii} doublet at 4481\AA, there is reasonable agreement
in the individual differential abundances, whilst the mean
value agree well with that for AV304.

\subsection{Silicon} 
In contrast with oxygen and magnesium, the differential silicon 
abundances show a wide spread, which is characterised by the
relatively large standard deviation in the mean value. The
reason for this is unclear as for most of the targets the
\ion{Si}{iii} lines near 4560\AA~are relatively well observed, 
although the relative strengths of these lines are sometimes
anomalous. Additionally for the cooler targets, the estimated 
abundances are very temperature sensitive; for example at an
effective temperature of 18\,000 K, a shift in temperature
of 1\,000 K changes the abundance by more than 0.3 dex.
However the \ion{O}{ii} ion has a similar temperature
sensitivity in this regime, whilst the oxygen abundance
estimates are far more consistent. Although the cause of the
discrepancies is unclear, they may well be due (at least in
part) to the small number of measurable silicon lines.
Indeed they are unlikely to reflect real abundance differences 
given the behaviour of other $\alpha$-process elements. 
The mean differential abundance is within the
uncertainties in good agreement with that found for
AV304.

\section{Discussion}
\subsection{The chemical composition of our B-type sample in NGC330}
The principle conclusion from the differential analysis is that
the cluster targets have a much higher nitrogen and possibly 
lower oxygen abundance than AV304. Relative nitrogen to oxygen 
abundances, $[\frac{N}{O}]$~of $-0.40\pm$0.24 dex and $-$1.6 dex  
are deduced for the NGC330 targets and AV304 respectively. For 
the former the estimates from the individual stars have been weighted 
by the number of lines observed (although if the stars are 
uniformly weighted the ratio is only changed by 0.1 dex), 
whilst for the latter the uncertainty in the individual element 
abundances imply an error in the ratio of approximately 0.2 dex.
For the NGC330 targets, if the errors in the ratios are randomly
distributed, the error on the mean would be reduced to approximately
$\pm$0.1 dex.  

As discussed in Paper I, the theoretical \ion{N}{ii} 
and \ion{O}{ii} line strengths have a similar dependence
on the adopted atmospheric parameters and hence the estimated
nitrogen to oxygen abundance ratios are unlikely to be affected
by uncertainties in these quantities. Hence we conclude that the 
$[\frac{N}{O}]$~ratio is enhanced by 1.2 dex with respect to AV304 
and this estimate is unlikely to be significantly affected by
uncertainties in atmospheric parameters or the relatively simple
LTE analysis adopted.

The simplest explanation for this nitrogen enhancement is that it
represents the products of hydrogen burning by the CNO bi-cycle.
In such circumstances, it might be expected that there were
a corresponding enhancement in helium and underabundances in
carbon and possibly oxygen, with the sum of CNO nuclei in the
NGC330 stars comparable to that in AV304.  It is therefore important
to try to map the abundances for the NGC330 targets as 
derived from our differential analysis onto an absolute scale.  
There are a number of options available to us, which we now discuss.

If we assume that the composition of AV304 is the baseline initial
composition of NGC330, then we can use the difference in the LTE 
abundances of the NGC330 targets relative to AV304. 
However we must also address the probable impact of non-LTE (NLTE) 
effects on these abundances. We have used NLTE calculations
similar to those discussed in McErlean et al. (\cite{McE00})
to estimate the difference in NLTE and LTE CNO abundance estimates
for AV304. The corrections are
approximately +0.07, -0.11 and -0.07 dex respectively for C, N 
and O.  In addition we note that our carbon abundance in AV304 relies
heavily on the 4267\AA\ line which is well known to
give spuriously low abundances. Following the discussion
of Vrancken et al. (\cite{Vra00}) and comparing their results
with those of Gies \& Lambert (\cite{Gie92}) we further
correct the carbon abundance by +0.34 dex.  Our final CNO
NLTE abundances estimates for AV304 are then 7.41, 6.55 and 8.16 dex 
in good agreement with the H\,{\sc ii} region results of 7.4, 6.6 
and 8.1 dex as summarized in the
discussion of baseline SMC abundances for A-type supergiants in the
SMC by Venn (\cite{Venn99}).  

We can now use these modified NLTE abundance estimates for AV304
and the difference in LTE abundances listed in Tables \ref{abs_analysis}
and \ref{gal_stand}
to estimate absolute CNO abundance for our NGC330 targets;
these are summarized in Table \ref{CNO}. In turn we can then estimate 
sum of the CNO and CN abundances for AV304, which are
are 8.24 and 7.46 dex respectively.  These may be compared with 
the mean NGC330 totals of 8.17 and 7.71 dex respectively.
There is some slight evidence for
an increase in CN but the difference of +0.25 should be compared
with uncertainties in the mean carbon and nitrogen abundances
of 0.15 and 0.18 dex (the total being dominated by the more
abundant species).  The sum of CNO is in good agreement
but again the total is dominated by the oxygen abundance
for which the uncertainty is approximately 0.13 dex.  
Clearly it is unproductive to compare summations of abundances
when one species is substantially
more abundant than all other species, and the uncertainty
in that abundance is similar to, or larger than, that of the 
less abundant species.  

We arrive at a similar picture if we instead consider the
differential abundances of the NGC330 stars relative
to BD+56\,576 using the NLTE abundances published by 
Vrancken et al. (\cite{Vra00}) to put them on an absolute
scale.  This results in CNO abundances for the NGC330 targets
of 7.28, 7.52 and 7.98 dex respectively which are in
good agreement with the values obtained using
AV304 as the standard (see Table \ref{CNO}).  This is possibly  
fortuitous, and given the difficulty in modeling
the 4267\AA\ line, unexpected for carbon. Nevertheless
it reinforces the previous discussion of the absolute
abundances.  We conclude that the chemical peculiarities
of the NGC330 targets may be understood in terms of nuclear 
processing by the CNO bi-cycle, perhaps with some weak
evidence that ON processing has occurred, but it is
not necessary to invoke primary nitrogen production
(although this cannot be precluded).

We can also search for correlations between element
abundances within the NGC330 targets. Linear least
squares fits show a positive correlation between the
helium and nitrogen abundances and negative correlations
between the carbon and nitrogen and between the oxygen
and nitrogen abundances. Interestingly all these trends
are consistent with the transformation of hydrogen into
helium using the CNO bicycle. Unfortunately however,
none of the correlations are convincing and the 
coefficients are not significantly different from
zero at even the 1$\sigma$\ level.

\subsection{Other stellar abundances in NGC330 and the SMC}

Korn et al. (\cite{Korn00}) have recently published  
C, O, Mg and Si NLTE abundances for another B-type giant
in NGC330, the star B30,  and their
abundances are in good agreement with ours given the 
magnitude of the NLTE corrections and the uncertainties
in both studies.  Unfortunately they do not give a nitrogen
abundance but two other similar SMC giants in their sample
have NLTE nitrogen abundances of 7.3 and 7.2 dex. We will 
return to these stars in the discussion of their evolutionary 
status below.  Furthermore an LTE nitrogen abundance for star B30 
was published by Reitermann et al. (\cite{Rei90}) who obtained 
a value of 7.4 dex for a microtubulence of 5 km/s. 

There have also been many studies of cool giants and 
supergiants in NGC330. These results are summarized in 
Table \ref{CNO}.  Of interest here are CNO abundances and
there are two recent estimates for NGC330 stars by 
Hill (\cite{Hil99}, H99) and Gonzalez \& Wallerstein (\cite{Gon99}, GW).
One obvious difference between these is that the 
mean nitrogen abundance of GW is systematically lower than that of H99.
However we note that the nitrogen abundance is derived from
molecular CN features, and depends on the adopted carbon
abundance (which is derived from C$_{2}$).  While H99 independently
derive their carbon abundances, GW adopted mean values
from the literature and there is a small systemic overestimation
relative to H99 (approximately 0.2 dex).  Such a small change
in the carbon abundance is the most likely reason for their
low nitrogen abundances and given that the carbon abundance of H99
may be more reliable and that their results agree better
with other samples of evolved B, A and F-type stars in the SMC, we
prefer their results for carbon and nitrogen.

Comparing with other SMC samples we note that our mean 
CNO abundances are in good agreement with the results
of Venn (\cite{Venn99}) and Hill et al. (\cite{Hil97})
although the NGC330 stars may be mildly metal poor.  
There have been previous suggestions that 
NGC330 may be relatively metal poor with respect to the SMC
(Grebel \& Richtler \cite{Gre92})
but our results confirm recent work in that any metal
deficiency must be small ($< 0.2$ dex).
We therefore conclude that the pattern of the mean CNO
abundances found in the NGC330 B-type giants is very similar to
that found for samples of other evolved A, F and K-type giants
and supergiants. We also note that Dufton et al. (\cite{Duf00})
investigated a large sample of B-type supergiants in the SMC
and the nitrogen abundances found in their less luminous stars,
although uncertain, are comparable to those found here.

\subsection{Evolutionary status}

\begin{figure*}
\epsfig{file=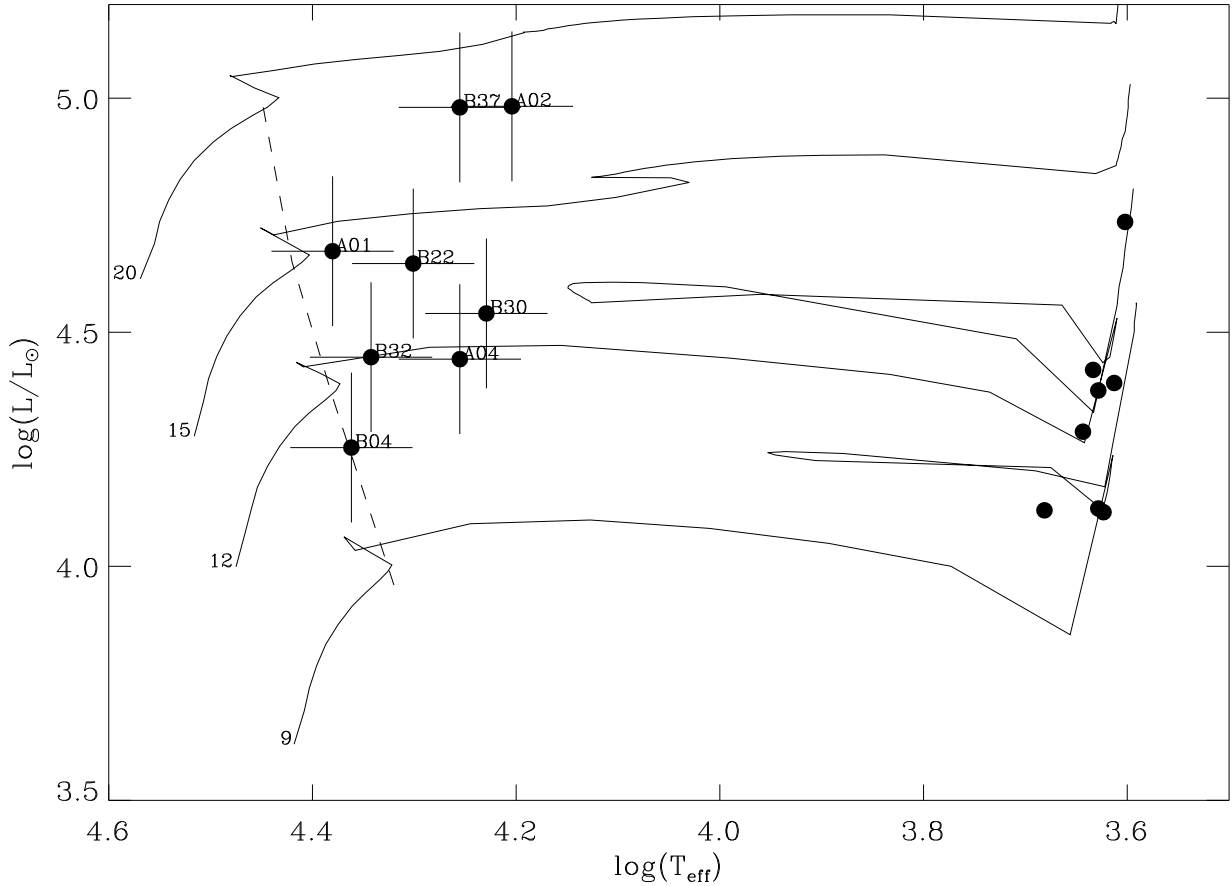}
\caption[]{HR-diagram for stars in NGC330 showing the positions of
the stars relative to the stellar evolution tracks (solid lines) of Charbonnel
et al. (\cite{Cha93}) which are computed for a metallicity of $Z = 0.004$.
Tracks are labeled with their initial masses. The dashed line represents
the approximate position of the end of the core H-burning main sequence
for the rotating models of Maeder \& Meynet (\cite{Mae01}) for an 
assumed initial rotational velocity of 300\,km/s.  Blue stars (analysed
here) are labeled with their identifications. For comparison we also
show the positions of the red supergiants in NGC330 analysed by Hill
(\cite{Hil99}) and Gonzalez \& Wallerstein (\cite{Gon99}), whose
chemical compositions we also discuss.  The error bars represent
typical uncertainties discussed in the text, for example 15\% in 
effective temperature.
}
\label{hrdiag}
\end{figure*}

It is useful to preface our discussion of the evolutionary
status of these stars by considering their positions
in the HR-diagram. We follow
the procedure described in VDLL, adopting a distance modulus
for the SMC of 18.9 and extinction estimates as discussed in
section 3.1.  As in VDLL we estimate spectroscopic masses
(M$_{spec}$) from our derived stellar radii and surface gravities.
Figure \ref{hrdiag} illustrates
the positions of the stars compared with the
non-rotating stellar evolution tracks of Charbonnel
et al. (\cite{Cha93}) which are computed with metallicity
of $Z = 0.004$ (SMC-like).  Again following VDLL we estimate
evolutionary masses (M$_{evol}$) by interpolation in this
diagram and these may be compared with M$_{spec}$ in Table \ref{masses}.
where all derived quantities are summarized.
We note that Maeder \& Meynet
(\cite{Mae01}) have produced a grid of calculations 
including the effects of rotation for a
metallicity of $ Z = 0.004$. Figure \ref{hrdiag} also shows the
locus of the end of H-burning main sequence phase for an assumed
initial rotational velocity of 300 km/s (taken from their
Figure 6). Note that this locus also corresponds to the
approximate position of the end of the main sequence for
the non-rotating models. This is a consequence of the
fact that the main sequence widening in Charbonnel et al. 
comes from their inclusion of convective overshooting, while 
a similar effect is obtained by Maeder \& Meynet with rotationally induced
mixing alone.  The fact that overshooting and
rotation both have similar results in this respect 
has been discussed previously, see for example 
Talon et al. (\cite{Tal97}).  We will return to this point later in the
discussion, after first considering the observed abundance pattern.

\begin{table*}
\caption{Derived quantities for B-type stars in NGC330: Stellar radii
(R/R$_{\odot}$), absolute visual magnitudes (M$_v$), bolometric 
corrections (B.C.), bolometric magnitudes (M$_{bol}$), luminosities
(log(L/L$_{\odot}$) and estimates of spectroscopic (M$_{spec}$)
and evolutionary (M$_{evol}$) masses in units of solar mass.
The distance modulus to the SMC was assumed to be 18.9 in the 
derivation of these quantities.  The last two columns compare
the difference in spectroscopic and evolutionary masses with
the uncertainties in log\,g ($\Delta$logg).
}
\label{masses}
\begin{center}
\begin{tabular}{lrllllrrrc}
\hline
\hline
Star & R/R$_{\odot}$ & M$_v$ & B.C. & M$_{bol}$ & log(L/L$_{\odot}$) 
& M$_{spec}$ & M$_{evol}$ & log(M$_{evol}$/M$_{spec}$) & $\Delta$logg \\
\hline
A01 & 12.6 & -4.48 & -2.50 & -6.98 & 4.67 & 29 & 15 & $-0.29$ & 0.3 \\
A02 & 40.4 & -6.28 & -1.49 & -7.77 & 4.98 & 12 & 18 & 0.18 & 0.1 \\
A04 & 17.2 & -4.64 & -1.77 & -6.41 & 4.44 &  7 & 12 & 0.23 & 0.2 \\
B04 &  8.5 & -3.58 & -2.34 & -5.92 & 4.25 & 10 & 11 & 0.04 & 0.2 \\
B22 & 17.6 & -4.89 & -2.03 & -6.92 & 4.65 & 11 & 14 & 0.10 & 0.2 \\
B32 & 11.5 & -4.17 & -2.23 & -6.40 & 4.45 &  5 & 12 & 0.38 & 0.3 \\
B37 & 31.9 & -5.99 & -1.76 & -7.75 & 4.98 &  9 & 18 & 0.30 & 0.1 \\
B30 & 21.6 & -5.08 & -1.64 & -6.72 & 4.54 & 11 & 13 & 0.07 & 0.2 \\
\hline
\end{tabular}
\end{center}
\end{table*}

Nitrogen enrichment is clearly a sign of contamination of
the surface by the products on the nuclear burning, in
this case from the CNO bi-cycle. What is especially interesting
about the current sample of stars (to which we can add 
the star B30 mentioned above) is that they represent a
coeval group of stars with rather homogeneous properties,
whose N/C surface abundance ratio is enhanced by typically
a factor of 10.  Stellar models which include the
effects of rotationally induced mixing have been proposed
as a means of producing the observed nitrogen enhancements
in main sequence and evolved massive stars.  Maeder \& Meynet
(\cite{Mae01}) have followed the evolution of the surface abundances 
for a range of initial rotational velocities.  Comparing 
our results to the models with high initial rotational
velocity, $v_{\rm ini} = 300$ km/s, and therefore relatively
large nitrogen enhancements, we find that the best agreement
is with their red supergiant or blue loop stars.  At the end
of the main sequence these models predict an increase in
[N/H] by a factor of only 3, or about 0.5 dex. Note however
that their initial abundance ratios are assumed to be solar and
therefore they overestimate the initial nitrogen abundance
significantly, adopting one fifth solar, which is
approximately 7.3 dex in our notation. 

We can perform a simple recalibration of their results by
assuming that a calculation with an initial lower nitrogen
abundance of 6.6 dex (but similar carbon and oxygen)
produces the same excess of nitrogen
in absolute terms. In this case the nitrogen abundance
at the end of the main sequence would be approximately
7.7 dex rather than 7.8 dex, a consequence of the fact that
the initial nitrogen abundance is negligible compared with
that which is produced by the star.  These models therefore
could conceivably reproduce the observed nitrogen
enhancements.  In fact if the initial N/O or N/(O+C)
abundance ratios are small and can be neglected, which in
the case of the SMC is a good approximation, then 
it is easy to show that one only needs a relatively 
small fraction of core material in ON equilibrium
to produce a big change in the observed nitrogen abundance.

Nevertheless, as Figure \ref{hrdiag} shows, 
there are other more serious
discrepancies. For example, our objects tend to lie 
red-wards of the main sequence despite the widening provided
by the rotating models (and models with convective overshooting). 
In addition all the B-type giants/supergiants
in our sample are slow rotators with values of $v$sin$i$ less than 
50\,km/s.  By contrast the rotating models require very 
high initial rotation to produce the enhanced nitrogen but do not 
predict significant slow down by the
end of the main sequence. As discussed in Paper I,
while there may be a selection effect in our observer
sample (we can only analyse the slow
rotators) it is highly unlikely that
there are so many fast rotators in NGC330 oriented pole on.
For example, both Mazzali et al. (\cite{Maz95})
and Keller \& Bessell (\cite{Kel98}) give $v$sin$i$ values for
22 of the brightest B and Be-type stars in NGC330.  They find that maximum
values lie in the range 300 -- 400\,km/s and if we assume that
our sample have similar rotation rates ($v$) this implies that
sin$i$ is less than 10 degrees.  In other words if the
distribution of $i$ is random we should expect our sample to be 
drawn from about 1.5\% of all the B-type stars in NGC330.  This is
clearly incompatible with the number of B-type stars in NGC330 in
the relevant magnitude range.
One is left with the conclusion that our objects are intrinsically
slow rotators at the present time.  Either they were
fast rotators in the past, and have somehow slowed down,
or some process other than rotationally induced mixing
is responsible for the observed abundance pattern.

Given the similarity between the carbon and nitrogen abundances
in the blue and red stars in NGC330 it is tempting to
invoke blue loops as a means of explaining our abundances.
This seems unlikely given that no stellar evolution calculations 
predict loops which progress hotter than effective temperatures
corresponding to late-B spectral types.  While Venn (\cite{Venn99})
invoked blue loops for the A-type supergiants this does
not seem a viable option for our early B-type stars, despite
similarities in CNO abundances.
Mass-transfer in binaries may also be invoked to explain
enhanced nitrogen abundances and Wellstein et al. (\cite{Wel01})
have recently produced models which produce nitrogen enriched
blue stars which can reside in the post main-sequence
gap. Such stars may also appear to be under-massive for their
luminosities and is therefore tempting to ascribe the 
discrepancies between spectroscopic and evolutionary masses
in Table \ref{masses} to binarity. One should be cautious
however because our spectroscopic masses were estimated
from the derived surface gravities and in some cases the 
uncertainties in this quantity are quite substantial.  The final
two columns of Table \ref{masses} compares the differences
between spectroscopic and evolutionary masses with the 
uncertainties in gravity and in all but three cases (A02, B32 and B37) the 
mass differences are easily accounted for.  There does appear
to be a suggestion of a correlation between nitrogen enhancement
and luminosity in our sample.   Stars A02 and B37 are the
most luminous and most N-rich of our sample,  however their
positions are perhaps consistent with being core-helium burning stars
on their way red-wards (this phase represented by the slight kink in
the 20 solar mass track for the non-rotating models).  
The problem for the rotating models
is that this phase becomes progressively cooler and shorter
lived as initial rotational velocity is increased, with only 
the slow rotators spending any significant time in this part 
of HR-diagram.  However, such stars should not be significantly
N-enriched, in contradiction to the observations.
B32 would appear to be the best candidate for the binary
evolution hypothesis, the real problem being that much
better constraints are needed on the gravity of this object, and
indeed all other stars in our sample, before definitive
statements can be made about possible mass discrepancies.
Finally, 
while binarity  appears to be an attractive scenario, it has a
significant problem in that
it is expected that the products of mass accretion will
be fast rotators having been spun up by the accreted material.
In addition radial velocities of all our stars are typical
of the NGC330 cluster (Table 1) although it must be noted that
expected radial velocity amplitudes of the kind of systems
predicted by the models of Wellstein \& Langer (\cite{Wel01})
are typically the order of 10--20 km/s.

Finally we note that VDLL carried out a similar study to the
present one but for the solar metallicity cluster
h+$\chi$ Per.  They did not find evidence for
significant nitrogen overabundances in any of the evolved B-type
stars in this cluster.  However if we take the nitrogen 
enrichments found here, in absolute terms, apply
these to their galactic counterparts it is clear that
this leads to enhancements the order of a factor
of 2--3.  For some objects in h+$\chi$ Per this
magnitude of a nitrogen enhancement is consistent
with the observations.  It is simply more obvious in the
SMC stars given their initial very low nitrogen abundance.

\begin{table*}
\caption{Comparison of our abundance estimates corrected for 
NLTE effects with results for other stars in NGC330, viz. the B-type
star B30 analysed by Korn et al. (NLTE C and O) and
Reitermann et al. (LTE N) (B30) and cool supergiants from
Hill (H99) and Gonzalez \& Wallerstein (GW). Also tabulated
are other SMC abundance estimates, viz. the main sequence star AV304
(Rolleston et al. \cite{Rol02} corrected for NLTE effects), 
A-type supergiants from 
Venn (A-stars), B-type giants and supergiants from Korn et al. 
(B-stars) and cool supergiants from Hill et al. (K-stars).}
\begin{center}
\label{CNO}
\begin{tabular}{llllllllll}
\hline
\hline
 & & \multicolumn{4}{c}{NGC330}& & \multicolumn{3}{c}{SMC supergiants} \\ 
\cline{3-6} \cline{8-10}
Element & AV304 & This paper 
& B30 & H99 & GW &
&A-stars & B-stars & K-stars \\ \hline
C & 7.47 & 7.26 & 7.3 & 7.40 & 7.55 & & -- & 7.40 & 7.55 \\
N & 6.55 & 7.52 & 7.4 & 7.21 & 6.96 & & 7.33 & 7.25 & 7.51 \\
O & 8.16 & 7.98 & 8.25 & 7.71 & 7.92 & & 8.14 & 8.15 & 8.01 \\ \hline
\end{tabular}
\end{center}
\end{table*}

\section{Summary}

In this paper we have presented new results for 5 B-type
giants/supergiants in the SMC cluster NGC330.  Together
with our previous work, plus results for one other such
star in NGC330 presented by Korn et al. this brings the
total of bright B-type stars analysed in this cluster to 8.  
All the stars have the following characteristics:
\begin{itemize}
\item They are nitrogen rich and possibly carbon poor, with perhaps
some marginal evidence for oxygen depletion, with their N/C ratios
being enhanced by approximately a factor of 10 relative to the 
SMC norm.  The nitrogen enhancements are remarkably homogeneous.
\item They all lie close to the main-sequence but with a
strong tendency to lie beyond the end of the main-sequence even
if main-sequence widening through rotation or convective
overshooting is considered.
\item They all appear to be currently slow rotators with $v$sin$i$ 
values below 50 km/s.  The sample is too large
for this to be an inclination angle effect for fast rotators.
\item The nitrogen enhancements agree well with those found for
cool red supergiants in the same cluster.
\end{itemize}

Single star stellar evolution models including the effects of 
rotation may indeed be able to explain the abundance patterns,
although we note that there is a clear need for stellar evolution
calculations with the correct mix of abundances for the
SMC, in particular scaling the solar composition 
produces much too high an initial nitrogen abundance for the
SMC.  However single 
star models suffer from severe difficulties in reproducing both
the effective temperatures of our sample and their low rotational
velocities.  Binary star models may be able to produce stars
in the correct effective temperature range but may also
have similar problems to single star
models in reproducing the low rotational velocities.
The similarity between blue B-type giant/supergiant nitrogen
enhancements and those found for red supergiants in NGC330
is surprising since one might expect the latter to be more
nitrogen enriched with the blue stars' enhancement coming from rotation
while that of the red stars coming from dredge-up processes.
In conclusion, the B-type stars considered here are still
something of a puzzle and not well modeled by current
stellar evolution calculations.  Moreover, given the large 
number of these objects in NGC330 it is clear that they
are not the product of some peculiar evolutionary
path and therefore represent an important challenge to
current stellar evolution theory.

\acknowledgements{Data reduction was performed on the PPARC funded
Northern Ireland {\sc starlink} node. DJL is grateful for NOVA
funding for a visit to Utrecht in May 2001 during which time
much of the present work carried out.  Thanks are also due
to a number of people for contributions to this project;
Norbert Langer, Paolo Mazzali, Gianni Marconi, 
Robert Rolleston and Kim Venn.}

\end{document}